# Effects of Turbulence Modeling and Parcel Approach on Dispersed Two-Phase Swirling Flow

Osama A. Marzouk*and E David Huckaby†

*Abstract*— Several numerical simulations of a co-axial particle-laden swirling air flow in a vertical circular pipe were performed. The air flow was modeled using the unsteady Favre-averaged Navier-Stokes equations. A Lagrangian model was used for the particle motion. The gas and particles are coupled through two-way momentum exchange. The results of the simulations using three versions of the $k-\epsilon$ turbulence model (standard, re-normalization group (RNG), and realizable) are compared with experimental mean velocity profiles. The standard model achieved the best overall performance. The realizable model was unable to satisfactorily predict the radial velocity; it is also the most computationally-expensive model. The simulations using the RNG model predicted additional recirculation zones. We also compared the particle and parcel approaches in solving the particle motion. In the latter, multiple similar particles are grouped in a single parcel, thereby reducing the amount of computation.

*Keywords: two-phase flow, swirl, turbulence model, particle, parcel*

## 1 Introduction

The use of computational fluid dynamics (CFD) to accurately model energy production systems is a challenging task [1]. Of current interest, due to every increasing energy demands, are coal-based energy systems such as pulverized coal (PC) boilers and gasifiers with an emphasis on systems which provide for carbon capture and storage (e.g. PC-oxyfuel). Turbulence and particle sub-models are one of many sub-models which are required to calculate the behavior of these gas-solid flow systems.

The particle-laden swirling flow experiment studied by Sommerfeld and Qiu [2] was selected as a test-case to assess the performance of three versions of the $k-\epsilon$ turbulence model for gas-solid flows. Previous numerical investigation of this experiment include, Euler-Lagrange (EL)/Reynolds-average Navier-Stokes (RANS-steady) [3], EL/large eddy simulations (LES) [4], Euler-Euler (EE)/RANS-unsteady [5], and EE/LES [6]. The extensive experimental measurements make this experiment a good test-case for gas-solid CFD.

A schematic of the experiment is shown in Fig. 1. The co-axial flow consists of a primary central jet, laden with particles at a loading of 0.034 kg-particles/kg-air and an annular secondary jet with a swirl number of 0.47 based on the inlet condition. Co-axial combustors have a similar configuration to this system. This is slightly below the nominal swirl number of typical burners which is greater than 0.6 [7]. The inlet swirl number was calculated as the ratio between the axial flux of angular momentum to the axial flux of linear momentum

$$S = \frac{2\int_0^{R_{sec}} \rho U_\theta U_x r^2 \, dr}{D_{cyl} \int_0^{R_{sec}} \rho U_x^2 \, r \, dr} \quad (1)$$

where $U_x$ and $U_\theta$ are the axial and tangential (swirl) velocities, $R_{sec} = 32$ mm is the outer radius of the swirling secondary jet, and $D_{cyl} = 197$ mm is internal diameter of cylinder into which the jets enter. The Reynolds number is approximately 52,400 (based on the outer diameter of the secondary jet). The particles were small spherical glass beads with a density of 2,500 kg/m$^3$, which were injected according to a log-normal distribution, with a mean number diameter of 45 $\mu$m.

In addition to investigating the effects of the turbulence modeling, we also study the effect of grouping similar particles in a parcel. In this approach, the equations of motion are integrated for each parcel, with a possibility of significant reduction in the computational resource required to simulate systems with a large number of particles (e.g., sprays).

## 2 Governing Equations

The continuity and momentum equations for the resolved density $\rho$, pressure $p$, and velocity $U_i$ fields are expressed and solved in the Cartesian coordinates as

$$\frac{\partial \rho}{\partial t} + \frac{\partial (\rho U_j)}{\partial x_j} = 0 \quad (2)$$

$$\frac{\partial (\rho U_i)}{\partial t} + \frac{\partial (\rho U_i U_j)}{\partial x_j} = -\frac{\partial p}{\partial x_i} + \frac{\partial (\sigma_{ij} + \tau_{ij})}{\partial x_j} + \rho g_i + S_p \quad (3)$$

*U.S. DOE, National Energy Technology Laboratory, 3610 Collins Ferry Road, P.O. Box 880, Morgantown, WV 26507-0880, Tel/Fax: 304-285-1396/0903, Email: omarzouk@vt.edu

†U.S. DOE, National Energy Technology Laboratory, 3610 Collins Ferry Road, P.O. Box 880, Morgantown, WV 26507-0880, Tel/Fax: 304-285-5457/0903, Email: E.David.Huckaby@NETL.DOE.gov

where $\sigma_{ij}$ and $\tau_{ij}$ are the viscous and Reynolds (turbulent) stress tensors, $g_i$ is gravitational vector (we only have $g_1 = 9.81\text{m/s}^2$), and $S_P$ is a source term accounting for the momentum from the particle-phase. As for Newtonian fluids, $\sigma_{ij}$ is calculated as

$$\sigma_{ij} = 2\mu\, S_{ij}^{dev}$$

where $\mu$ is the dynamic viscosity and $S_{ij}^{dev}$ is the deviatoric (traceless) part of the strain-rate tensor $S_{ij}$

$$S_{ij} = \frac{1}{2}\left(\frac{\partial U_i}{\partial x_j} + \frac{\partial U_j}{\partial x_i}\right)$$

$$S_{ij}^{dev} = \frac{1}{2}\left(\frac{\partial U_i}{\partial x_j} + \frac{\partial U_j}{\partial x_i} - \frac{2}{3}\delta_{ij}\frac{\partial U_k}{\partial x_k}\right)$$

The tensor $\tau_{ij}$ is not resolved directly. Instead, its effects are represented using the gradient transport hypothesis

$$\tau_{ij} = 2\mu_t\left(S_{ij}^{dev}\right) - \frac{2}{3}\rho k\, \delta_{ij} \quad (4)$$

This brings a new variable, namely the turbulent (or eddy) viscosity $\mu_t$. Different eddy-viscosity turbulence models propose different strategies to calculate $\mu_t$. In the case of $k - \epsilon$ models, it is calculated as

$$\mu_t = C_\mu \rho \frac{k^2}{\epsilon} \quad (5)$$

which brings two new variables, namely $k$ and $\epsilon$ (the turbulent kinetic energy per unit mass and its dissipation rate). They are obtained by solving two coupled transport equations. The form of these equations varies depending on the model implementation. The *standard* $k - \epsilon$ model refers to the Jones-Launder form [8], without wall damping functions, and with the empirical constants given by Launder and Sharma [9]. We consider here three implementations, which are described in the following subsections. The wall-function approach is used to model the near wall behavior for all three turbulence models.

### 2.1 Standard $k - \epsilon$ Model

We start with the standard version, with the $k$ and $\epsilon$ equations are

$$\frac{\partial(\rho k)}{\partial t} + \frac{\partial(\rho U_j k)}{\partial x_j} = \frac{\partial}{\partial x_j}\left[\left(\mu + \frac{\mu_t}{\sigma_k}\right)\frac{\partial k}{\partial x_j}\right] + P - \rho\epsilon \quad (6)$$

$$\frac{\partial(\rho\epsilon)}{\partial t} + \frac{\partial(\rho U_j \epsilon)}{\partial x_j} = \frac{\partial}{\partial x_j}\left[\left(\mu + \frac{\mu_t}{\sigma_\epsilon}\right)\frac{\partial \epsilon}{\partial x_j}\right]$$
$$+ \frac{\epsilon}{k}\left(C_{\epsilon 1} G - C_{\epsilon 2}\rho\epsilon\right) - \left(\frac{2}{3}C_{\epsilon 1} + C_{\epsilon 3}\right)\rho\epsilon\frac{\partial U_k}{\partial x_k} \quad (7)$$

where $P$ is the production of kinetic energy due to the gradients in the resolved velocity field

$$P = \tau_{ij}\frac{\partial U_i}{\partial x_j}$$

which is evaluated as

$$P = G - \frac{2}{3}\rho k \frac{\partial U_k}{\partial x_k}$$

with

$$G = 2\mu_t\, S_{ij}^{dev}\frac{\partial U_i}{\partial x_j} = 2\mu_t\left(S_{ij}S_{ij} - \frac{1}{3}\left[\frac{\partial U_k}{\partial x_k}\right]^2\right)$$

The addition of $C_{\epsilon 3}$ in the last term on the rhs of (7) is not in the standard model. It was proposed [10, 11] for compressible turbulence. However, we will refer to this implementation as the standard model. The model constants are

$C_\mu = 0.09$, $\sigma_k = 1.0$, $\sigma_\epsilon = 1.3$, $C_{\epsilon 1} = 1.44$,
$C_{\epsilon 2} = 1.92$, $C_{\epsilon 3} = -0.33$,

We note that a constant named $C_{\epsilon 3}$ also appears in another version of the $k - \epsilon$ model as part of a buoyancy-induced turbulence generation term. The temperature gradients are not significant in the current problem and we do not consider buoyancy effects on turbulence.

### 2.2 Re-Normalization Group $k - \epsilon$ Model

The RNG model was developed [12, 13] using techniques from re-normalization group theory. The $k$ and $\epsilon$ equations have the same form in (6) and (7), but the constants have different values. In addition, the constant $C_{\epsilon 1}$ is replaced by $C_{\epsilon 1}^*$, which is no longer a constant, but is determined from an auxiliary function as

$$C_{\epsilon 1}^* = C_{\epsilon 1} - \frac{\eta(1 - \eta/\eta_0)}{1 + \beta\eta^3}$$

where

$$\eta = \frac{k}{\epsilon}\sqrt{2\, S_{ij}\, S_{ij}}$$

The model constants are

$C_\mu = 0.0845$, $\sigma_k = 0.7194$, $\sigma_\epsilon = 0.7194$, $C_{\epsilon 1} = 1.42$,
$C_{\epsilon 2} = 1.68$, $C_{\epsilon 3} = -0.33$, $\eta_0 = 4.38$, $\beta = 0.012$

We should mention here that another version (e.g., in Ref. [14]) of the RNG model replaces the constant $C_{\epsilon 2}$ by a function $C_{\epsilon 2}^*$ while keeping $C_{\epsilon 1}$ constant at 1.42. Both versions have been used in different studies and were referred to as RNG $k - \epsilon$ model. The version considered here was used, for example, in a recent study [15] that involves not only turbulence, but also combustion, soot formation, and radiation.

### 2.3 Realizable $k - \epsilon$ Model

The realizable $k - \epsilon$ model was formulated [16] such that the calculated normal Reynolds stresses are positive definite and off diagonal Reynolds shear (off-diagonal) stresses satisfy the Schwarz inequality. Similar to the RNG model, the form of the $k$ equation is the same as the one in (6). In addition to altering the model constants, the two main modifications lie in replacing the constant $C_\mu$ used in calculating the eddy viscosity in (5) by a function, and in changing the rhs (the production and destruction terms) of the $\epsilon$ equation. The last term in (7) is dropped. With this, the $\epsilon$ equation becomes

$$\frac{\partial(\rho\epsilon)}{\partial t} + \frac{\partial(\rho U_j \epsilon)}{\partial x_j} = \frac{\partial}{\partial x_j}\left[\left(\mu + \frac{\mu_t}{\sigma_\epsilon}\right)\frac{\partial \epsilon}{\partial x_j}\right]$$
$$+ C_1 \rho \epsilon S - C_{\epsilon 2}\rho \frac{\epsilon^2}{k + \sqrt{(\mu/\rho)\epsilon}} \qquad (8)$$

where

$$C_1 = \max\left(0.43, \frac{\eta}{\eta + 5}\right)$$

$$\eta = \frac{k}{\epsilon}\sqrt{2 S_{ij} S_{ij}} \quad \text{(as in the RNG model)}$$

$$C_\mu = \frac{1}{A_0 + A_S(U^* k/\epsilon)}$$

$$U^* = \sqrt{S_{ij}S_{ij} + \Omega_{ij}\Omega_{ij}}$$

$\Omega_{ij}$ is rate-of-rotation tensor $= \frac{1}{2}\left(\frac{\partial U_i}{\partial x_j} - \frac{\partial U_j}{\partial x_i}\right)$

$$A_S = \sqrt{6}\cos(\phi); \; \phi = \frac{1}{3}\arccos\left(\sqrt{6}\,W\right)$$

$$W = \min\left[\max\left(2\sqrt{2}\,\frac{S_{ij}S_{jk}S_{ik}}{S^3}, -\frac{1}{\sqrt{6}}\right), \frac{1}{\sqrt{6}}\right]$$

The model constants are

$$\sigma_k = 1.0, \; \sigma_\epsilon = 1.2, \; C_{\epsilon 2} = 1.9, \; A_0 = 4.0$$

## 2.4 Wall Function

In wall-bounded flows, a very thin viscous (laminar) sublayer exists near the wall. Whereas it is possible to solve the flow equations all the way to the wall (including the viscous sublayer), this requires modifications in the turbulence models described in the previous subsections because they are based on fully-turbulent (high Reynolds number) flows. In addition, resolving the flow in this thin region with appropriate resolution (at least five cells within the steep-gradient viscous sublayer [17]) requires very fine meshes that incur an extensive amount of calculations per time step.

The high-Reynolds number version of the turbulence models are examined here and thus the wall-function treatment is required. This approach avoids the solution of the governing equations of the flow inside the viscous sublayer by utilizing empirical laws which relate the wall conditions to values of the dependent variables just outside the viscous sublayer. In the current implementation (related to the collective work in Refs. [18, 19, 20]), the $\epsilon$ equation is not solved and an algebraic expression is used instead. The term $G$ in the production term $P$ in the $k$ equation is evaluated from another expression, but the $k$ equation is still solved. Finally, the turbulent viscosity is evaluated from an expression other than the one in (5), which does not require $\epsilon$. The treatment is applied at the first cell node next to a wall. If this node is found (estimated) to lie within the viscous sublayer, $\mu_t$ is set to zero, which is consistent with the physics of the problem.

The subsequent expressions describe the wall-function treatment as implemented in the 1.5 version of the finite volume open source code OpenFOAM [21, 22] (open field operation and manipulation). This was used to perform all the simulations here. The coordinate normal to a wall is denoted by $y$. In addition, the subscript $P$ refers to values at the cell node adjacent to a wall. The wall-function treatment requires an auxiliary nondimensional variable $y_P^+$, which is a measure of the normal distance from the wall, and is calculated (using old $k_P$ from the available $k$ field) as

$$y_P^+ = \frac{C_\mu^{1/4}\sqrt{k_P}\,y_P}{\mu_P/\rho_P} \qquad (9)$$

The turbulent viscosity at $y_P$ is evaluated from

$$(\mu_t)_P = \mu_P \begin{cases} 0 & , \; y_P^+ \leq y_{Lam}^+ \\ \frac{y_P^+ \kappa}{\ln(E\,y_P^+)} - 1 & , \; y_P^+ > y_{Lam}^+ \end{cases} \qquad (10)$$

where $E = 9.0$ is a nondimensional constant and $\kappa = 0.4187$ is the von Kármán constant. These two parameters are used to calculate the interface of the viscous sublayer and the log-layer, $y_{Lam}^+$, through the following iterative formula:

$$y_{Lam}^+ = \frac{\ln\left(E\,y_{Lam}^+\right)}{\kappa} \qquad (11)$$

which converges to 10.967.

The production of kinetic energy in a cell adjacent to a wall, $G_P$, is evaluated using

$$G_P = \begin{cases} 0 & , \; y_P^+ \leq y_{Lam}^+ \\ \frac{C_\mu^{1/4}\sqrt{k_P}}{\kappa\,y_P}[(\mu_t)_P + \mu_P]\frac{|U_P|}{y_P} & , \; y_P^+ > y_{Lam}^+ \end{cases} \qquad (12)$$

where $|U_P|/y_P$ approximates $\partial|U_{tangent}|/\partial y$ at $y_P$. The turbulence dissipation rate in the cell adjacent to a wall, $\epsilon_P$, is evaluated from

$$\epsilon_P = \frac{C_\mu^{3/4}k_P^{3/2}}{\kappa\,y_P} \qquad (13)$$

This expression is based on the assumption of local turbulence equilibrium, $P = \epsilon$ (as in the log-layer), giving [14, 23]

$$\epsilon_P = \frac{u_\tau^3}{\kappa\,y_P} \qquad (14)$$

where $u_\tau$ is the friction velocity, which is formally defined as $\sqrt{|\tau_w|/\rho_w}$. However, under the assumption of local turbulence equilibrium, $u_\tau$ is evaluated from

$$u_\tau = C_\mu^{1/4}\sqrt{k} \quad (15)$$

The expression in (13) follows from (14) and (15). In fact, (15) was used to eliminate $u_\tau$ from (9) and (12).

## 3 Particle Motion

The Lagrangian equations of motion of a particle are

$$\frac{d\mathbf{x}}{dt} = \mathbf{u} \quad (16a)$$

$$m\frac{d\mathbf{u}}{dt} = \mathbf{f} \quad (16b)$$

where $m$ is the constant mass of the particle, $\mathbf{u}$ is the particle velocity, and $\mathbf{f}$ is the force acting on the particle. In this study; the drag, gravity, and buoyancy are considered, thus the force $\mathbf{f}$ has the following form [24]:

$$\mathbf{f} = -\frac{\pi d^2}{8}\rho C_D |\mathbf{u}-\mathbf{U}^*|(\mathbf{u}-\mathbf{U}^*) + m\mathbf{g} - \rho\forall\mathbf{g} \quad (17)$$

where $d$ is the particle diameter, $C_D$ is the drag coefficient (which is a function of the particle Reynolds number, $\mathrm{Re}_d$ as will be described later), $\forall$ is the particle volume and $\mathbf{U}^*$ is the instantaneous fluid velocity,

$$\mathbf{U}^* \equiv \mathbf{U} + \mathbf{U}' \quad (18)$$

The vector $\mathbf{U}$ is the resolved velocity of the fluid (interpolated at the particle location) which is calculated after solving the governing equations of the flow, coupled with the turbulence model. The fluctuating velocity, $\mathbf{U}'$, is estimated using the discrete random walk algorithm [25, 26]. In this algorithm, uncorrelated eddies are generated randomly, but the particle trajectory is deterministic within the eddy. The fluctuating velocity affects the particle over an interaction time, $T_{interac}$, which is the minimum of the eddy life time (Lagrangian integral time scale of turbulence), $T_{eddy}$, and the residence or transit time, $T_{cross}$. The latter is the time needed by the particle to traverse the eddy. These characteristic times are calculated as

$$T_{eddy} = \frac{k}{\epsilon} \quad (19a)$$

$$T_{cross} = C_{cross}\frac{k^{3/2}}{\epsilon|\mathbf{u}-\mathbf{U}-\mathbf{U}'|} \quad (19b)$$

$$T_{interac} = \min(T_{eddy}, T_{cross}) \quad (19c)$$

In (19b), $\mathbf{U}'$ is lagged from the previous time step, and $C_{cross} = 0.16432$. This values is equal to $C_\mu^{3/4}$ (where the standard value of $C_\mu = 0.09$ is used regardless of the implemented turbulence model). The turbulence information, thus the characteristic times in (19), are updated every time step to account for the fact that the turbulence encountered by the particle in its trajectory is not homogeneous.

The drag coefficient for a sphere is determined from the following two-region formula [10], which is very similar to the Schiller-Naumann [27] expression for $\mathrm{Re}_d \leq 1000$, and uses a constant Newton drag coefficient for $\mathrm{Re}_d > 1000$

$$C_D = \begin{cases} \frac{24}{\mathrm{Re}_d}\left(1+\frac{1}{6}\mathrm{Re}_d^{2/3}\right), & \mathrm{Re}_d \leq 1,000 \\ 0.424, & \mathrm{Re}_d > 1,000 \end{cases} \quad (20)$$

where the particle Reynolds number is defined as

$$\mathrm{Re}_d = \frac{\rho|\mathbf{u}-\mathbf{U}^*|d}{\mu} \quad (21)$$

Combining (16) and (17), and using $m = \rho_s \pi d^3/6$ ($\rho_s$ is the 'solid' particle density), the particle's equations of motion become

$$\frac{d\mathbf{x}}{dt} = \mathbf{u} \quad (22a)$$

$$\frac{d\mathbf{u}}{dt} = -\frac{\mathbf{u}-\mathbf{U}^*}{\tau} + \left(1-\frac{\rho}{\rho_s}\right)\mathbf{g} \quad (22b)$$

where

$$\tau = \frac{4}{3}\frac{\rho_s d}{\rho C_D|\mathbf{u}-\mathbf{U}^*|} = \frac{\rho_s d^2}{18\mu}\frac{24}{\mathrm{Re}_d C_D}$$

is the nondimensional momentum relaxation time of the solid particle.

The particle position is tracked using the algorithm described by Macpherson et al. [28]. The algorithm consists of a series of substeps in which the particle position, $\mathbf{x}$, is tracked within a cell using a forward (explicit) Euler scheme followed by integration of the particle momentum equation using a backward (implicit) Euler scheme to update the particle velocity, $\mathbf{u}$. When calculating the resultant force due to the particle on the fluid phase, the algorithm takes into account the particle residence time in each cell. Interaction with the wall is represented through elastic collisions. The tangential friction with the wall is neglected.

## 4 Mesh and Boundary Conditions

The problem is treated as axisymmetric (although the results are mirrored in some figures for better visualization). The domain starts at the expansion location with $x = 0$ in Fig. 1 and extends to $x = 1.0$ m. The domain is a 3D wedge (opening angle $5^o$), with a front area of 1.0 m×0.097 m, and 240 and 182 mesh points in the axial and radial directions. The mesh is nonuniform both axially and radially, with finer resolution near walls and between the two jets. The mesh has 40,080 cells. The inlet condition for the velocity in the primary (inner) jet is specified in terms of the mass flow rate (9.9 g/s). For the secondary (outer) jet, the inlet velocity is specified using the experimental velocity profile. A zero-gradient condition is applied to the pressure at the inflow. The turbulent kinetic energy, $k$, is set to 0.211 m$^2$/s$^2$ and 0.567 m$^2$/s$^2$ in the primary and secondary jets, respectively; and the dissipation rate, $\epsilon$, is set to 0.796 m$^2$/s$^3$ and 3.51 m$^2$/s$^3$ in the primary and secondary jets, respectively. The inflow $k$ was estimated assuming 3% turbulence intensity (the experimental value was not specified, but 3% is a reasonable medium-turbulence level [29]) and the inflow $\epsilon$ was then estimated from [19, 30]

$$\epsilon = C_\mu^{3/4}\, k^{1.5}/l \qquad (23)$$

where the standard value 0.09 is used for $C_\mu$, and $l$ is the turbulence length scale, which is approximated as ≈10% of the cylinder diameter ($l = 0.02$ m). At the outflow, zero-gradient conditions are applied for all variables except the pressure, where a constant value of $10^5$ N/m$^2$ is imposed. At the walls, the wall-function treatment is used for the turbulence, and a zero-gradient condition is used for the pressure.

The PISO (pressure implicit splitting of operators) scheme was used to solve the governing flow equations. A variable time step is adjusted dynamically to limit the maximum CFL to 0.3. The backward Euler scheme is used for the time integration of the flow equations. Upwind differencing is used for the convective terms. Linear (second-order central difference) interpolation is used to find the mass fluxes at the face centers from the nodal values, and is also used for the diffusion terms.

The particle mass flow rate is 0.34 g/s, which corresponds to 0.00472 g/s for our case of $5^o$ wedge. The parcels are injected at a speed of 12.5 m/s, which is the nominal axial inflow velocity in the primary jet. The current particle injection model implemented in OpenFOAM does not allow one to specify both a *constant* mass injection rate and *constant particles-per-parcel*, (ppp) due to the distribution of the particles diameter (mass), which is sampled from a log-normal PDF. Therefore, if the mass flow rate of particles is fixed, then ppp can be below or above the target value. On the other hand if ppp is fixed then instantaneous particle mass injection will vary about the specified mean.

## 5 Results

The simulated flow time is 0.6s for the results presented in this paper. We have found that this time interval is sufficient for all particles to traverse the domain and to achieve a stationary flow in the gas-phase. The last 0.1s of this interval is used to obtain the mean gas-phase velocities.

Figure 2 shows three snapshots of the parcels after 0.05s, 0.1s, and 0.15s using the standard $k - \epsilon$ model (the diameters of the parcels are evenly scaled by a factor of 100). This figure illustrates that the model captures well the expected dynamics of the particles. The larger particles (with larger inertia) maintain their axial motion, penetrating the central recirculation bubble, and are not affected strongly by the swirl and radial velocity of the gas-phase. Smaller particles are entrained due to smaller relaxation times and are directed to the walls.

The mean axial, radial, and tangential velocities of the gas-phase are shown in Fig. 3. The negative mean axial velocity along the centerline and the walls identify the regions of recirculation. The strong variations in all velocities are confined to a distance of 150 mm after the inlet. Axial and tangential velocities exhibit an initial decay, whereas the radial velocity increases at the upstream boundary of the central recirculation bubble.

A comparison between the mean streamlines obtained with the three turbulence models is given in Fig. 4. Besides the central bubble, we can see secondary recirculation zones (due to the sudden expansion) at the corners located at the top of the cylinder. The standard model gives the shortest recirculation bubble, with best agreement with the experimental results. The realizable model gives a longer bubble, but with a qualitatively similar structure. The RNG model resolves, in addition to the central and two secondary recirculation zones, two noticeable tertiary recirculation zones at the beginning (top) of the central bubble. This feature was not reported in the experimental results.

The mean velocity fields are sampled at several different axial stations. We compare the mean velocity components from the three turbulence models with the measured values at 4 stations in Figs. 5−8, located at $x = 3$ mm, 52 mm, 112 mm, and 155 mm. The first two stations are located upstream of the central bubble, whereas the last two span the central bubble. At $x = 3$ mm, all models predict results for the axial and tangential velocities profiles that are in agreement with the measured profiles. The radial velocity using the realizable model shows considerable disparity, with excessively negative (inward) velocity in the region away from the jets, followed by an outward flow near the wall, which is opposite to the inward flow observed experimentally. The standard model has a slightly better agreement with the measurements than the RNG model at this axial location. At $x = 52$ mm, the standard and realizable models behave similarly except for the radial velocity, where the realizable model overpredicts the radial velocity in the region $r > 70$ mm. The RNG model gives a higher axial velocity in the region $r < 20$ mm than the other models, which is closer to the measurements. Unfortunately, this is accompanied by underprediction of the radial velocity for $r < 30$ mm. At $x = 112$ mm, the RNG predictions deviate considerably from the measurements. This is a direct consequence of the tertiary recirculation shown in Fig. 4. As in the earlier stations, the standard and realizable models provide similar results except for the radial velocity, with the realizable model failing to capture the measured peak at $r \approx 80$ mm. The standard model provides better prediction of the axial velocity in the vicinity of the wall than the other two models. At $x = 155$ mm, the realizable model underpredicts the axial velocity near the centerline. This can be explained by the longer central bubble. The standard model provides better agreement with the measurements. The standard and realizable models provide similar predictions for the tangential velocity, which match well with the measurements for $r < 50$ mm. The RNG model underpredicts this velocity. The standard model shows the best agreement of the three modes for the radial velocity, and the RNG model provides the poorest prediction.

On a computing machine with two processors: quad core Intel Xeon L5335 2.00GHz, a simulation interval of 0.5s required 17.13hr CPU time for the standard model, 19.44hr for the RNG model, and 24.81hr for the realizable model. The standard model has the lowest computational demand due to its relative simplicity. The realizable model is the most computationally-expensive version, with CPU time equal to 145% and 128% the CPU times in the case of the standard and RNG models, respectively.

The standard $k - \epsilon$ model was selected to examine the effect of the number of particles-per-parcel, *ppp*. This version of the model was selected, based on the results of the previous subsection, in which the standard turbulence model was demonstrated to have the best overall performance. The three selected values of *ppp* are: 1, 10, and 100. The first value corresponds to the particle approach, where each particle is tracked individually. Large values of *ppp* may adversely affect the resolution of the particle-phase and as a consequence, the accuracy of the simulations. In addition, the point-force (also called point-mass) treatment, used here for the particles, requires that the parcel is smaller than the cell in which it is located. This puts an additional limitation on the *ppp* parameter. In Fig. 9, three snapshots of the parcels are shown at $t = 0.2$s. The parcels are scaled by the particle diameter times factors of 100, 215, and 464 for *ppp* = 1, 10, and 100; respectively. These factors correspond to $\sqrt[3]{ppp}$ times an arbitrary constant of 100 to provide consistent scaling of the parcels. As shown in the figure, the case with *ppp* =100 has lost some of the particle-phase characteristics, in terms of shortened convection of the heavier parcels and weakened entrainment (outside the central bubble) of the smaller parcels. Therefore, we suspect that this case is representative with regard to particle motion. However, it is a good test-case serving our objective of examining the sensitivity of the mean gas-phase field to *ppp*. To this end, we considered the axial profiles of the mean axial velocity at the centerline (middle of the primary jet) and the radial and tangential velocities at $r = 25$ mm (middle of the secondary annular jet). The profiles are given in Figs. 10 and 11, respectively. There is no discernible effect of using different *ppp* values.

Lastly, we compare the computational time for the three simulated cases (for 0.5s flow time), where the difference is a measure of the computational saving by reducing the number of parcels equations to be solved. For *ppp*=1 and 100, the CPU times are 31.15hr and 29.84hr. The corresponding numbers of injected parcels over this simulation interval are 12,099 and 143. At the end of the simulation interval, there were 12,017 and 142 parcels in the domain, respectively. These simulations were conducted on identical machines (with two processors: dual core AMD Opteron 265 1.8GHz). Therefore, the computing-time saving is only 4.2% when *ppp* is increased by two orders of magnitude. This indicates that most of the computing time is spent in solving the gas-phase equations as a result of the low particle loading.

## 6 Conclusions and Future Work

We simulated a co-axial air flow with a particle-laden primary jet and a swirling annular jet, entering a sudden expansion. Three versions of the $k - \epsilon$ model: standard, RNG, and realizable were applied. The standard model had the best overall performance based on the mean gas-phase velocities. The RNG model showed considerable deviations from the measurements in some regions. The main drawback of the realizable model is its erroneous prediction of the radial velocity. The primary differences in the predicted velocity profiles were related to the different flow structures and mean streamlines at the central recirculation bubble. The mean gas-phase velocity field was not sensitive to whether a particle or parcel approach is used. However, this can be due to the low particle loading considered here. Further studies at higher particle loadings would help better identify the relationship between accuracy and the number of particles-per-parcel.

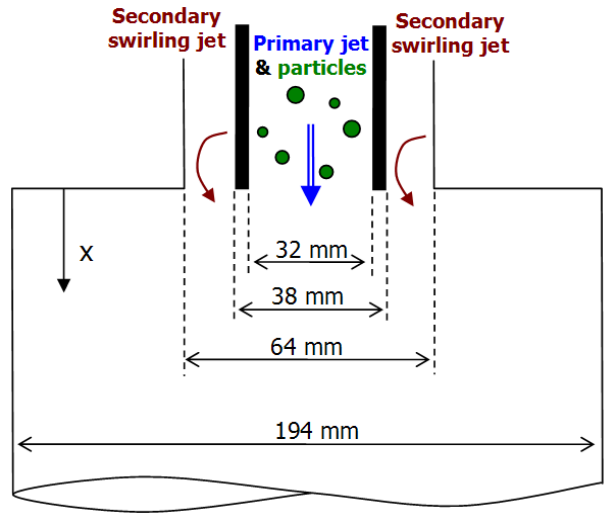

Figure 1: Illustration of the jet flow.

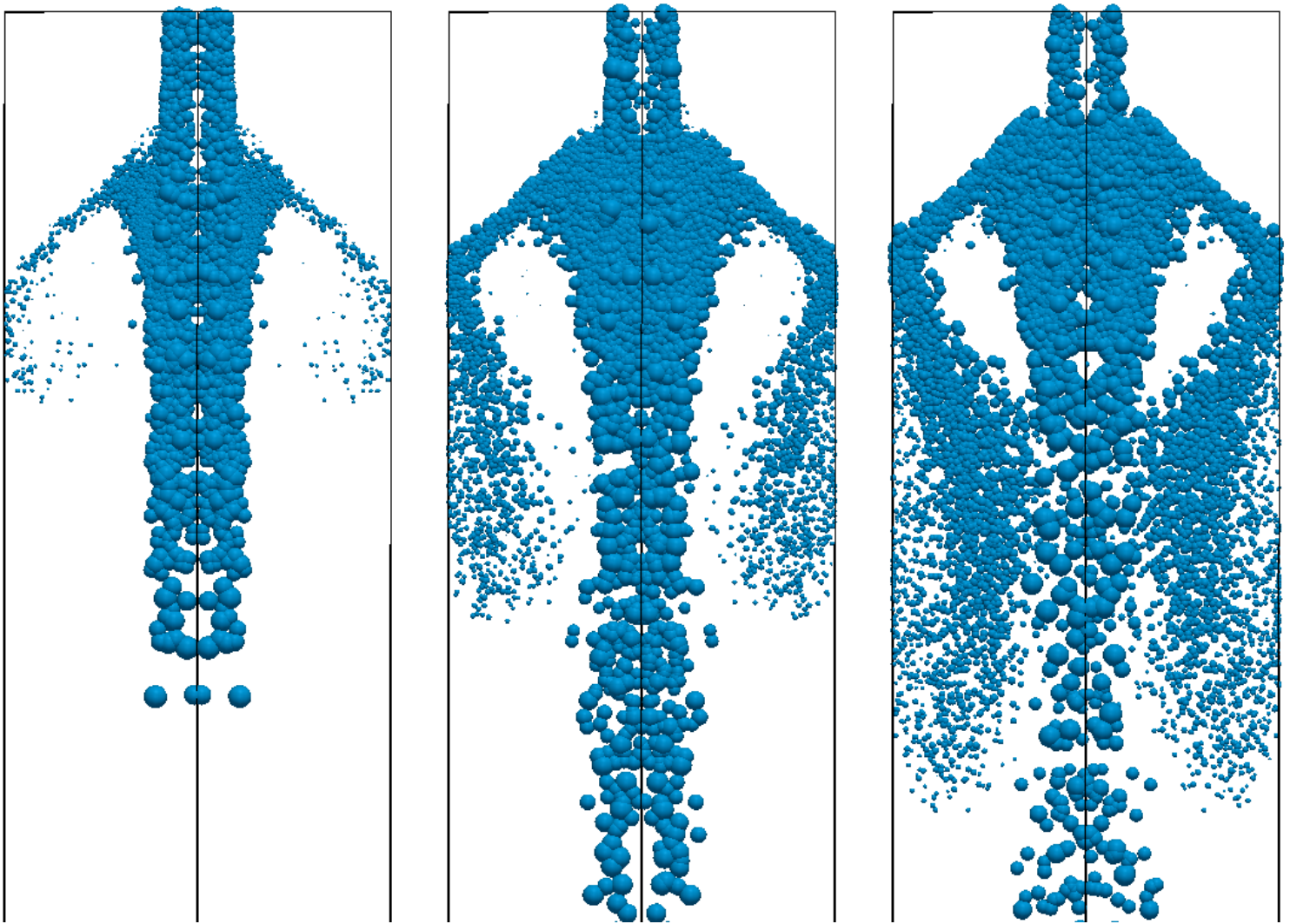

Figure 2: Parcel motion using the standard $k - \epsilon$ model at $t = 0.05$s (left), $t = 0.10$s (middle), and $t = 0.15$s (right).

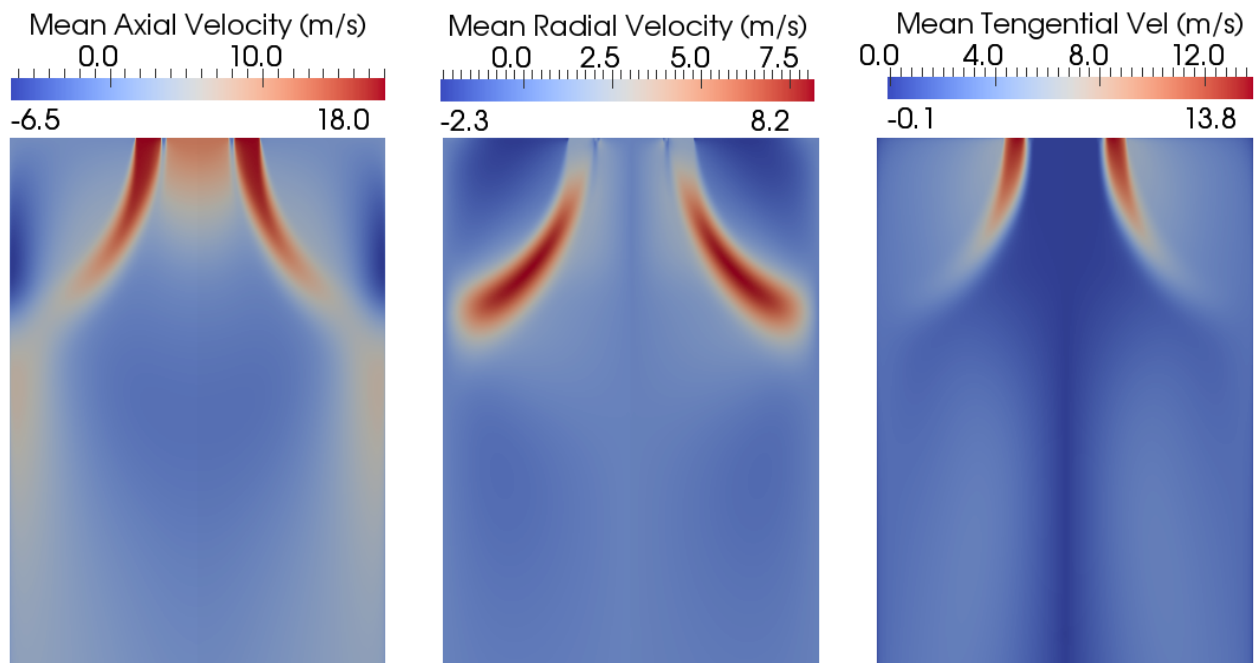

Figure 3: Mean gas-phase velocity components near the inlet using the standard $k - \epsilon$ model.

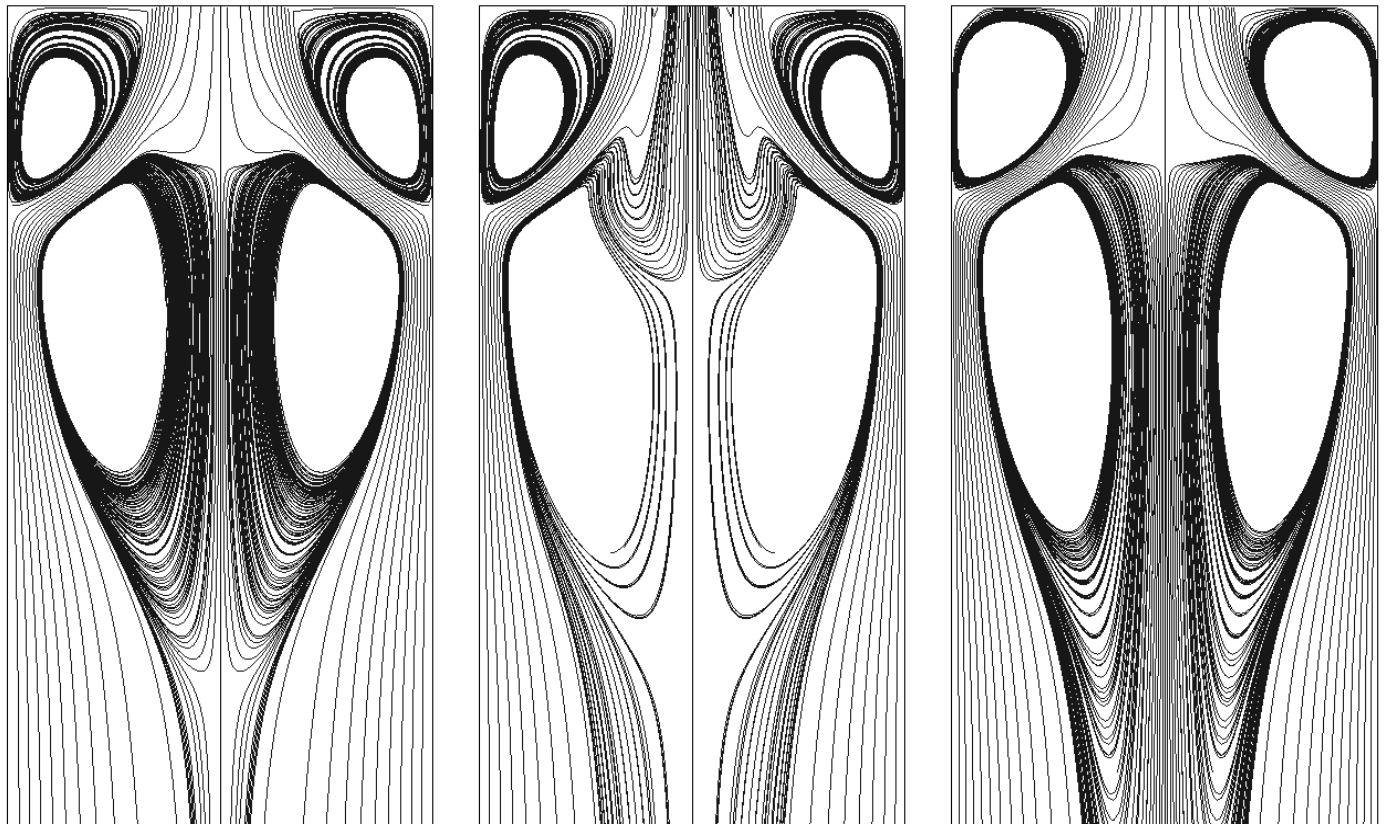

Figure 4: Comparison of the mean streamlines with 3 $k - \epsilon$ models: standard (left), RNG (middle), and realizable (right).

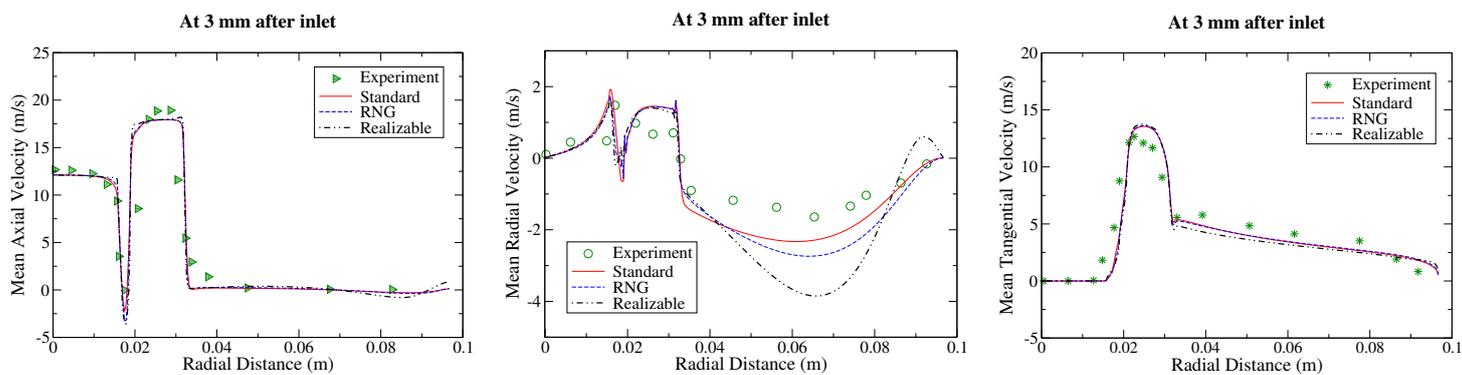

Figure 5: Comparison of the mean velocity components at $x = 3$ mm with three $k - \epsilon$ turbulence models.

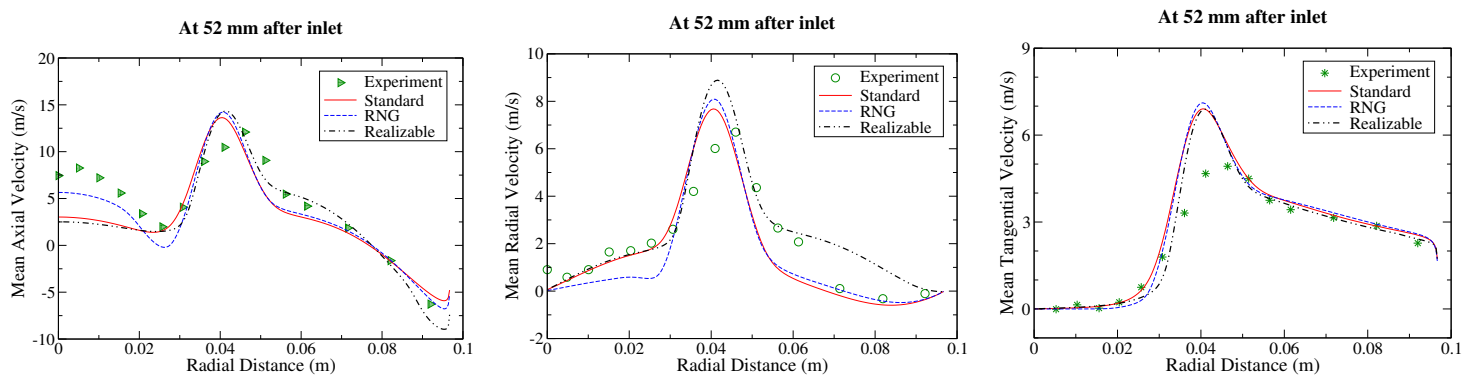

Figure 6: Comparison of the mean velocity components at $x = 52$ mm with three $k - \epsilon$ turbulence models.

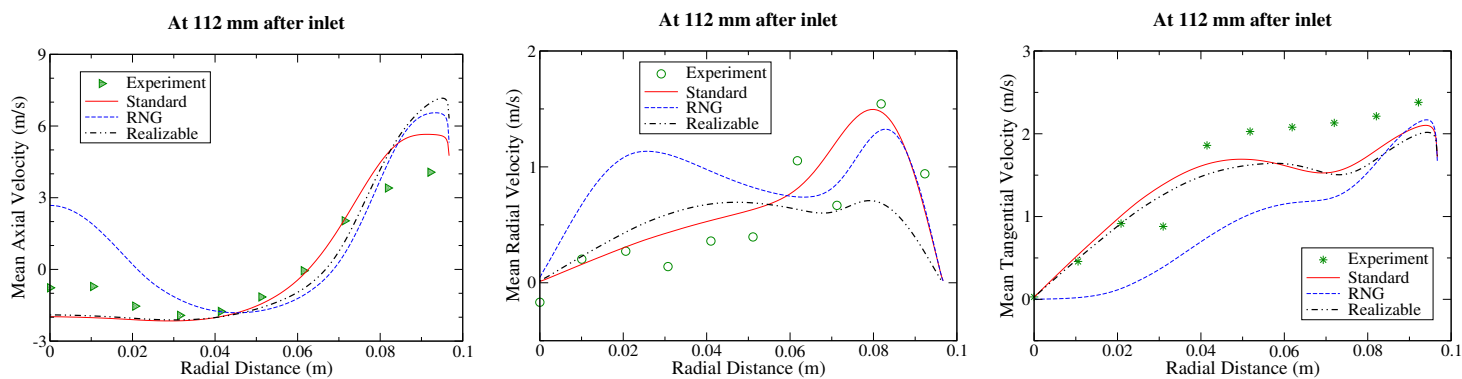

Figure 7: Comparison of the mean velocity components at $x = 112$ mm with three $k - \epsilon$ turbulence models.

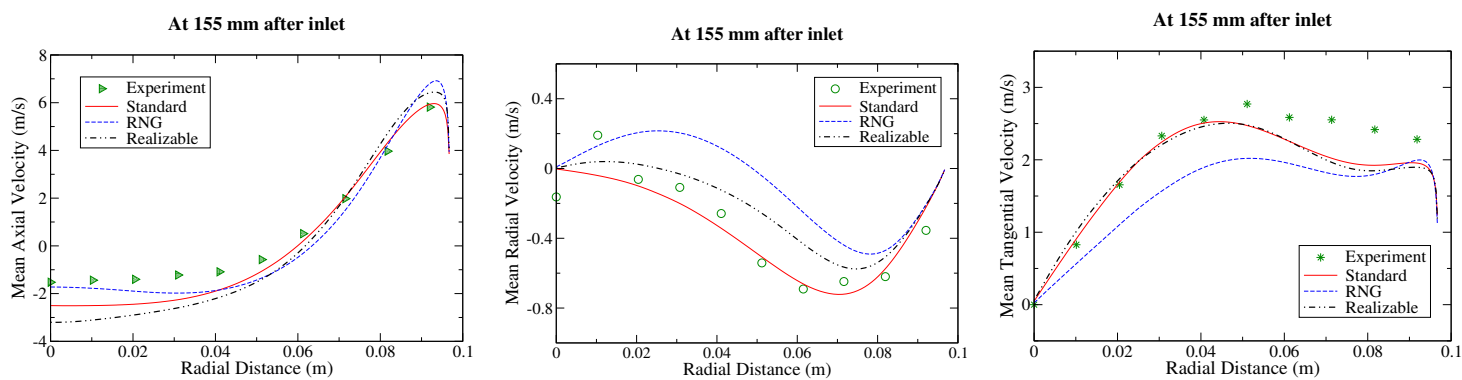

Figure 8: Comparison of the mean velocity components at $x = 155$ mm with three $k - \epsilon$ turbulence models.

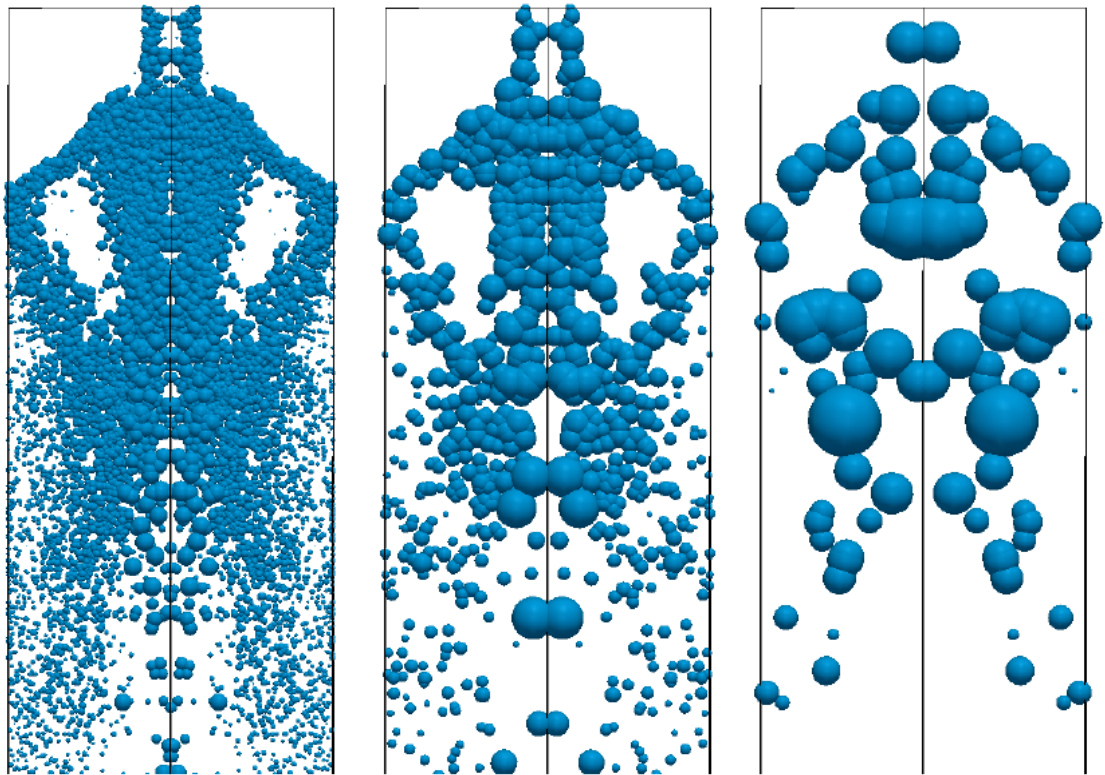

Figure 9: Comparison of parcels motion at $t = 0.2$s with different values of particles-per-parcel, *ppp*: 1 (left), 10 (middle), and 100 (right).

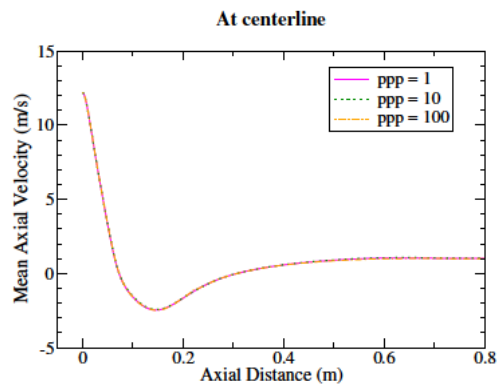

Figure 10: Comparison of the mean axial velocity with different particles-per-parcel at the centerline.

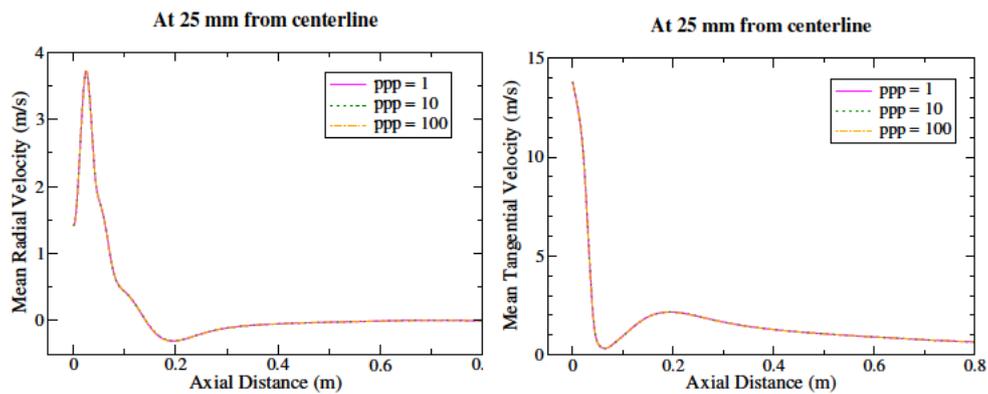

Figure 11: Comparison of the mean radial and tangential velocities with different particles-per-parcel at $r = 25$ mm.